\documentstyle[pre,aps,epsbox]{revtex}

\tighten

\begin{document}

\draft
\title{
Cluster Property and Robustness of\\
Ground States
of Interacting Many Bosons
\footnote{
to appear in J.\ Phys.\ Soc.\ Jpn.\ Vol.\ 71 No.\ 1 (2002)
}
}
\author{
Akira {\sc Shimizu}
\footnote{E-mail: shmz@ASone.c.u-tokyo.ac.jp}
 and Takayuki {\sc Miyadera}
}
\address{
Department of Basic Science, University of Tokyo, 
Komaba, Meguro-ku, Tokyo 153-8902, Japan
}
\maketitle
\begin{abstract}
e study spatial correlation functions of local operators
of interacting many bosons confined 
in a box of a large, but {\em finite} volume $V$, 
for various `ground states' 
whose energy densities are almost degenerate.
The ground states include
the coherent state of interacting bosons (CSIB),
the number state of interacting bosons (NSIB), 
and 
the number-phase squeezed state of interacting bosons,
which interpolates between the CSIB and NSIB.
It was shown previously that only the CSIB is robust
(i.e., decoheres much more slowly than the other states)
against the leakage of bosons into an environment.
We show that for the CSIB 
the spatial correlation 
of any local operators $A({\bf r})$ and $B({\bf r'})$
(which are localized around ${\bf r}$ and ${\bf r'}$, respectively)
vanishes 
as $|{\bf r} - {\bf r'} | \sim V^{1/3} \to \infty$, i.e., 
the CSIB has the `cluster property.'
In contrast, the other ground states do not possess 
the cluster property.
Therefore, 
we have successfully shown % for interacting many bosons 
that the robust state has the cluster property.
% whereas fragile states do not.
This ensures the consistency of the field theory of bosons 
with macroscopic theories.
\end{abstract}
\bigskip


\bigskip

%%%%%%% Introduction %%%%%%%%

It is {\em required} % and usually {\em assumed} 
that microscopic 
theories should be consistent with % macroscopic experiences and 
macroscopic theories such as thermodynamics.
However, it has not been {\em shown}  
whether existing microscopic theories of real physical systems
are indeed consistent with macroscopic theories.
This paper examines one of the consistency conditions\cite{eg}, 
and show that
it is indeed satisfied
by a standard microscopic theory of interacting many bosons.

We consider a quantum system of a macroscopic volume $V$.
When the system is closed and $V \to \infty$, 
the cluster property is one of the most fundamental properties
of pure states (pure phases) \cite{ruelle,haag}.
Here, a state $\omega$ is said to have the cluster property\cite{odlro}
iff 
\begin{equation}
\left|
\omega(\hat A({\bf r}) \hat B({\bf r'}))
- \omega(\hat A({\bf r})) \omega(\hat B({\bf r'}))
\right|
\to
0
\quad
{\rm as} \ 
| {\bf r} - {\bf r'} | \to \infty
\label{CP}\end{equation}
for {\em any} local operators
$\hat A({\bf r})$ and $\hat B({\bf r'})$
(at an equal time \cite{equal})
which are localized around ${\bf r}$ and ${\bf r'}$, respectively.
Here, $\omega(\cdot)$ denotes the expectation 
value in the state $\omega$.
Any pure states of infinite systems 
should have the cluster property \cite{ruelle,haag}.
This ensures, for example, 
that 
fluctuations of intensive variables are negligible 
\cite{ruelle,HL,miyashita,KT}, 
in consistency with thermodynamics.
This consistency should be generalized to the case of {\em finite} $V$, 
because thermodynamics is applicable to finite systems as well.
In quantum theories of finite systems, 
one can construct various pure states in a single Hilbert space.
However, they are not necessarily consistent with 
thermodynamics: 
some states exhibit anomalously large fluctuations of 
intensive variables \cite{ruelle,KT,HL,miyashita}. 
Nevertheless, as long as $V < + \infty$, 
such anomalous states are allowed 
as pure states in quantum theory, according to the 
uniqueness of representation for finite systems \cite{ruelle,haag}.
(If one simply takes the limit of
$V \to \infty$, such states approach states which 
do not possess the cluster property, 
hence do not remain pure states \cite{ruelle,haag,HL,miyashita,KT}.)
Therefore, to be consistent with thermodynamics, 
there should be some mechanism(s) which excludes anomalous states
from pure states in finite systems.

To explore this point, we note that % in these discussions 
the cluster property and fluctuations are defined only 
as {\em static} properties of {\em closed} systems.
However,
any finite systems of macroscopic sizes 
(except for the whole universe)
would be interacting with their surrounding environments.
Hence, one must also consider 
{\em dynamical} properties of {\em open} systems.
In general, the environments induce decoherence, 
and pure states would evolve into mixed states as time 
evolves.
When some states 
decohere much faster than the other states, 
we say such states are {\it fragile}.
When, on the other hand, some states 
decohere much slower than the other states,
we say such states are {\it robust}.
Examples of fragile and robust states were given, e.g., in refs.
\cite{zurek81,barnett,SMprl2000}, 
where it was suggested that 
% Some state 
% may decohere much faster than the other states.
% We say such a state is {\it fragile}.\cite{zurek81,barnett,SMprl2000}
% Another state 
% may decohere much slower than the other states.
% We say such a state is {\it robust}.\cite{zurek81,barnett,SMprl2000}
fragile states would be much difficult to 
prepare and observe than robust states.\cite{fragile}

These observations lead to the following conjecture:
it is necessary for a microscopic theory to be consistent with 
macroscopic theories that
{\em states with anomalous fluctuations are fragile}
(hence difficult to prepare and observe), 
whereas {\em robust states are 
non-fluctuating states} 
(which possess the cluster property in the limit of 
$V \to \infty$).
Although this very fundamental requirement has been suggested
in many places (e.g., in refs.\ \cite{ruelle,KT}), 
% it was reserved as an assumption so far. Namely, 
it has not been known yet (to the authors' knowledge)
whether general quantum field theories indeed satisfy this requirement.
[Conversely, if this is not the case, 
the fundamental requirement may place a limit on possible
forms of field theories in such a way that the requirement is 
satisfied.]
Even for a particular model or theory of a macroscopic system, 
it was not examined yet whether the requirement is satisfied.
From a microscopic point of view, it is non-trivial --- may be 
surprising --- that 
a {\em dynamical} property (robustness) 
of an {\em open} system subject to dissipations
is directly related to  
a {\em static} property  (cluster property) of a {\em closed} system.

In this paper, we show that the fundamental requirement is 
indeed satisfied 
by a standard microscopic theory in condensed matter, i.e., 
the field theory of bosons with repulsive interaction.
For interacting many bosons confined in a large but {\em finite} box,
many wave functions 
had been proposed as the `ground states,'
whose energy densities are almost degenerate
(completely degenerate when $V \to \infty$).\cite{SMprl2000,EandC}
Among them are, e.g., 
Bogoliubov's ground state, and 
the ground state that has {\em exactly} $N$ bosons,
which we call the number state of interacting bosons (NSIB).\cite{SI}
In a previous paper \cite{SMprl2000}, 
we already studied the robustness, against leakage of bosons
into an environment, 
of these wave functions.
It was shown there that
most of them are fragile, and that a robust state is 
an exceptional wave function which we call 
the coherent state of interacting bosons (CSIB).
Hence, 
we shall show that 
the fundamental requirement is satisfied
by showing that 
the CSIB possesses the cluster property
(in the limit of $V \to \infty$), 
whereas the other ground states do not.

%%%%%%% Preliminary %%%%%%%%%%

For bosons confined in a finite box of volume $V$ with the 
periodic boundary conditions, the CSIB with amplitude 
$\alpha$ ($= |\alpha| e^{i \varphi}$) is defined by
\cite{SMprl2000,SI}
\begin{equation}
| \alpha, {\rm G} \rangle
\equiv
e^{- |\alpha|^2/2}
\sum_{N=0}^{\infty}
\frac{\alpha^N}{\sqrt{N!}}
|N, {\rm G} \rangle,
\label{CSIB}\end{equation}
where
$
|N, {\rm G} \rangle
$ denotes the NSIB.
Note that $N$ is well-defined because $V$ is finite, 
and thus the vector $|N, {\rm G} \rangle$ exists.
The explicit form of $|N, {\rm G} \rangle$ was given in refs. \cite{SI,GA} 
in the case of 
weakly-interacting bosons.
Since  $|N, {\rm G} \rangle$ has a complicated wave function, 
the CSIB is totally different from 
the coherent state of free bosons, which is defined by
$ % \begin{equation}
| \alpha, 0 \rangle
\equiv
e^{- |\alpha|^2/2}
\sum_{N=0}^{\infty}
(\alpha^N / \sqrt{N!})
|N, 0 \rangle,
$ % \label{CSFB}\end{equation}
where
$
|N, 0 \rangle
$ denotes the number state of {\em free} bosons, 
in which $N$ bosons occupy the 
lowest single-body state whereas the other single-body states are
vacant.
On the other hand, 
some properties of the CSIB are similar to those of 
the coherent state of free bosons; 
e.g., 
$\langle N \rangle = \langle \delta N^2 \rangle =|\alpha|^2$,
hence $\langle \delta N^2 \rangle / \langle N \rangle = 1$.
Since we are interested in macroscopic systems,
for which $\langle N \rangle$ is macroscopically large, 
we assume $|\alpha| \gg 1$ in the following analysis.

Before going further, a few words are worth mentioning here:
(i) The above wave function of the CSIB might look against a
superselection rule, which `forbids' coherent 
superpositions of states with different numbers of (massive)
bosons. 
However, we previously showed that such superpositions
are allowed for a subsystem of a huge system
if the wave function of the total system is 
appropriately taken\cite{SMprl2000,ssr}.
(ii) The robust state $|\alpha, {\rm G} \rangle$ is not an
energy eigenstate (whereas 
an energy eigenstate $|N, {\rm G} \rangle$ is fragile).
Hence, 
even when interactions with the environments are negligible, 
the CSIB (prepared at $t=0$) evolves with time
(in the Schr\"odinger picture).
Putting
$\alpha(t) \equiv e^{i (\varphi - \mu t/\hbar)} | \alpha|$, 
$\mu \equiv (\partial E_{N, {\rm G}}/\partial N )_{N=|\alpha|^2}$
(with $E_{N, {\rm G}}$ being the eigenenergy of $|N, {\rm G} \rangle$),
and ${\cal G} \equiv (E_{N,G} - \mu N)_{N=|\alpha|^2}$,
we find the time evolution as
\begin{equation}
| \alpha, {\rm G}; t \rangle
=
%e^{-i(E_{|\alpha|^2,G} - \mu |\alpha|^2)t/\hbar}
e^{-i {\cal G} t/\hbar}
\ |
\alpha(t), {\rm G}
\rangle,
\label{CSIBt}\end{equation}
where terms of order $1/V$ have been neglected.
These neglected terms 
cause a {\em spontaneous} collapse of the wave function \cite{wright,EandC}, 
i.e., a collapse through the {\em internal} dynamics
(without perturbations from an environment).
The collapse time $t_{\rm coll}$ increases with 
$V$ (e.g., $t_{\rm coll} \propto V^{1/2}$ for a uniform 
system\cite{wright,EandC}), 
and becomes of a macroscopic time scale 
for a large $V$. 
In this work, 
{\em we are interested in a much shorter time range}
$0 \leq t \ll t_{\rm coll}$.
% where $t_{\rm interest} = {\cal O}(V^0)$.
% (We can easily take this range to include 
% the decoherence time of fragile states decohere if the system is perturbed 
% by an environment 
% (In the case of ref.\ \cite{SMprl2000}, for example, 
% $t_{\rm dec}^{\rm fragile} \simeq 1/\langle N \rangle j(n) \ll t_{\rm coll}$, 
% where $j(n)$ is given by eq.\ (9) of ref.\ \cite{SMprl2000}.)
Hence, the ${\cal O}(1/V)$ terms can be neglected in 
the derivation of eq.\ (\ref{CSIBt}),
and the CSIB does not collapse spontaneously; 
the time evolution only induces the phase rotations
of $\alpha$. 
(The overall phase factor $e^{-i {\cal G} t/\hbar}$ has no physical meaning.)
Therefore, it is sufficient to examine the cluster property at $t=0$ for 
a general value of $\alpha$ ($|\alpha| \gg 1$).
This should be contrasted with the coherent state of free bosons: 
it collapses spontaneously due to 
boson-boson interactions, and $t_{\rm coll} = {\cal O}(V^0)$.
As a result, % we can show that 
although 
it has the cluster property at $t=0$, 
the property will be lost when $t = {\cal O}(V^0)$.
In this paper, 
we are not interested in such states, and
we will only consider the states
(such as the CSIB and NSIB)
which are stable enough
(i.e., $t_{\rm coll}$ is much longer than ${\cal O}(V^0)$) 
against the internal dynamics. 

To examine the cluster property, 
we decompose the boson field $\hat \psi$ into 
the anomalous and regular parts \cite{SMprl2000,SI,LP}:
\begin{equation}
\hat \psi
=
\hat \Xi + \hat \psi'.
\label{decomposition}\end{equation}
Here, $\hat \Xi$ is the anomalous part, which connects
the ground states of different numbers of bosons as
\begin{equation}
\hat \Xi | N, {\rm G} \rangle
=
\sqrt{N} \xi | N-1, {\rm G} \rangle,
\end{equation}
where
$ %\begin{equation}
\xi 
\equiv
\langle N-1, {\rm G}| \hat \psi | N, {\rm G} \rangle/\sqrt{N}.
$ % \end{equation}
The magnitudes of $\xi$ characterizes the condensation,\cite{Ndep} and 
we here consider the condensed states for which 
\begin{equation}
\sqrt{N} \xi = {\cal O}(1).
\label{condense}\end{equation}
Note that the CSIB is an eigenstate of $\hat \Xi$, i.e., 
$
\hat \Xi 
| \alpha, {\rm G} \rangle
=
\alpha \xi | \alpha, {\rm G} \rangle
$ \cite{SMprl2000,SI}.
On the other hand, the regular part 
$\hat \psi'$ transforms $| N, {\rm G} \rangle$ into 
(a superposition of)
excited states (of $N-1$ bosons):
\begin{equation}
{\langle N - \Delta N, {\rm G} |}  \hat \psi' {| N, {\rm G} \rangle}
=
0,
\ \mbox{hence} \
\langle \beta, {\rm G}| \hat \psi' | \alpha, {\rm G} \rangle
=
0.
\label{pda}\end{equation}
Note however that 
$
\hat \psi' | N, {\rm G} \rangle
\neq 0
$
and
$
\hat \psi' | \alpha, {\rm G} \rangle
\neq 0
$.
For example, 
\begin{equation}
\int
\langle \alpha, {\rm G}|
 \hat \psi^{\prime \dagger} \hat \psi' 
| \alpha, {\rm G} \rangle
d^3 r
=
\langle N \rangle - \langle N_0 \rangle
\neq 0,
\end{equation}
where $\langle N_0 \rangle$ is 
the so-called ``number of the condensate particles" \cite{SI,LP}.
For weakly-interacting bosons, we can show by explicit calculations that
\begin{equation}
\langle N, \nu |
[\hat \Xi, \hat \Xi^\dagger]
| N', \nu' \rangle
, \
\langle N, \nu |
[\hat \Xi, \hat \psi']
| N', \nu' \rangle
, \
\langle N, \nu |
[\hat \Xi, \hat \psi^{\prime \dagger}]
| N', \nu' \rangle
\ = s(1/V),
\label{com_rel}\end{equation}
where 
$
| N, \nu \rangle
$
and
$
| N', \nu' \rangle
$
are energy eigenstates that have exactly $N$ bosons 
(here, $\nu$ and $\nu'$ label them; e.g., 
$| N, \nu \rangle = | N, G \rangle$ for $\nu = {\rm G}$), and
$s(x)$ denotes a smooth function that vanishes as
$x \to 0$.
Lifshitz and Pitaevskii \cite{LP}
claimed eq.\ (\ref{com_rel}) even for bosons with stronger interactions.
If this is the case, 
the following results are applicable to a wide range of interaction 
strength.

%%%%% Proof of the CP %%%%%%%%

We first examine the case where the local operators in 
eq.\ (\ref{CP}) take the 
following forms:
\begin{eqnarray}
\hat A({\bf r})
&=&
\hat \psi^\dagger ({\bf r})
\hat \psi ({\bf r})
\hat \psi^\dagger ({\bf r})
\cdots,
\label{A}
\\
\hat B({\bf r})
&=&
\hat \psi^\dagger ({\bf r})
\hat \psi^\dagger ({\bf r})
\hat \psi ({\bf r})
\cdots.
\label{B}
\end{eqnarray}
Their correlation for the CSIB
% for $|{\bf r} - {\bf r'}| \sim V^{1/3}$, 
is evaluated as
\begin{eqnarray}
&&
\langle \alpha, {\rm G} | 
\hat A({\bf r})
\hat B({\bf r'})
| \alpha, {\rm G} \rangle  
\nonumber \\
&& =
\langle \alpha, {\rm G} | 
\{ \hat \Xi^\dagger + \hat \psi^{\prime \dagger} ({\bf r}) \}
\{ \hat \Xi + \hat \psi' ({\bf r}) \}
% \{ \hat \Xi^\dagger + \hat \psi^{\prime \dagger} ({\bf r}) \}
\cdots
\{ \hat \Xi^\dagger + \hat \psi^{\prime \dagger}  ({\bf r'}) \}
\{ \hat \Xi^\dagger + \hat \psi^{\prime \dagger}  ({\bf r'}) \}
% \{ \hat \Xi + \hat \psi' ({\bf r'}) \}
\cdots
| \alpha, {\rm G} \rangle  
\nonumber \\
&& =
\langle \alpha, {\rm G} | 
\mbox{
(all $\hat \Xi^\dagger$'s are moved to the left, 
and all $\hat \Xi$'s are moved to the right)
}
| \alpha, {\rm G} \rangle
+ s(1/V)
\nonumber \\
&& =
\langle \alpha, {\rm G} | 
\mbox{
(all $\hat \Xi^\dagger$'s are replaced with $\alpha^* \xi^*$, 
and all $\hat \Xi$'s are replaced with $\alpha \xi$)
}
| \alpha, {\rm G} \rangle
+ s(1/V)
\nonumber \\
&& =
\langle \alpha, {\rm G} | 
\{ \alpha^* \xi^* + \hat \psi^{\prime \dagger} ({\bf r}) \}
\{ \alpha \xi + \hat \psi'({\bf r}) \}
% \{ \Xi^* + \hat \psi^{\prime \dagger} ({\bf r}) \}
\cdots
\{ \alpha^* \xi^* + \hat \psi^{\prime \dagger} ({\bf r'}) \}
\{ \alpha^* \xi^* + \hat \psi^{\prime \dagger} ({\bf r'}) \}
% \{ \Xi + \hat \psi'({\bf r'}) \}
\cdots
| \alpha, {\rm G} \rangle  
+ s(1/V),
\label{tochuu}\end{eqnarray}
where use has been made of eq.\ (\ref{com_rel}).
As mentioned above (eq.\ (\ref{pda})), % (eqs.\ (\ref{pdN}) and (\ref{pda})), 
$\hat \psi'$ and $\hat \psi^{\prime \dagger}$ 
transform the ground states into excited states.
Since any excitation cannot propagate a long distance in zero time
interval, equal-time correlations of $\hat \psi'$ vanish
for a large $|{\bf r} - {\bf r'}|$.
(This fact has been used, e.g.,  
in the standard argument on the ODLRO \cite{LP}.)
Hence, the last line of eq.\ (\ref{tochuu}) approaches, 
as $|{\bf r} - {\bf r'}| \sim V^{1/3}$, 
\begin{eqnarray}
& &
\langle \alpha, {\rm G} | 
\{ \alpha^* \xi^* + \hat \psi^{\prime \dagger} ({\bf r}) \}
\{ \alpha \xi + \hat \psi'({\bf r}) \}
% \{ \Xi^* + \hat \psi^{\prime \dagger} ({\bf r}) \}
\cdots
| \alpha, {\rm G} \rangle  \langle \alpha, {\rm G} | 
\{ \alpha^* \xi^* + \hat \psi^{\prime \dagger} ({\bf r'}) \}
\{ \alpha^* \xi^* + \hat \psi^{\prime \dagger} ({\bf r'}) \}
% \{ \Xi + \hat \psi'({\bf r'}) \}
\cdots
| \alpha, {\rm G} \rangle  
+ s(1/V)
\nonumber \\
&& =
\langle \alpha, {\rm G} | 
\{ \hat \Xi^\dagger + \hat \psi^{\prime \dagger} ({\bf r}) \}
\{ \hat \Xi + \hat \psi' ({\bf r}) \}
% \{ \hat \Xi^\dagger + \hat \psi^{\prime \dagger} ({\bf r}) \}
\cdots
| \alpha, {\rm G} \rangle  \langle \alpha, {\rm G} | 
\{ \hat \Xi^\dagger + \hat \psi^{\prime \dagger}  ({\bf r'}) \}
\{ \hat \Xi^\dagger + \hat \psi^{\prime \dagger}  ({\bf r'}) \}
% \{ \hat \Xi + \hat \psi' ({\bf r'}) \}
\cdots
| \alpha, {\rm G} \rangle  
+ s(1/V).
\label{calcAB}\end{eqnarray}
Therefore, 
\begin{equation}
\langle \alpha, {\rm G} | 
\hat A({\bf r})
\hat B({\bf r'})
| \alpha, {\rm G} \rangle  
=
\langle \alpha, {\rm G} | 
\hat A({\bf r})
| \alpha, {\rm G} \rangle  \langle \alpha, {\rm G} | 
\hat B({\bf r'})
| \alpha, {\rm G} \rangle  
+ s(1/V)
\quad
\mbox{
for $|{\bf r} - {\bf r'}| \sim V^{1/3}$}.
\label{AB}\end{equation}
By similar calculations,
we also obtain the relation (\ref{AB}) 
for other forms of $\hat A$ and $\hat B$, including 
those which consist of derivatives of the field operators.
Therefore, eq.\ (\ref{CP}) is 
indeed satisfied in the limit of $V \to \infty$
(while keeping the boson density $\langle N \rangle /V$ constant)
by the CSIB, for any local operators.

%%%% More rigorously %%%%%%%

More rigorously, we must perform smoothing of 
the field operators \cite{ruelle,haag}
because, for example, $\hat \psi^\dagger({\bf r}) | 0 \rangle$ 
cannot be a vector in a Hilbert space since
its norm diverges like $\lim_{{\bf r} \to {\bf 0}} \delta({\bf r})$,
whereas any non-zero vector in a Hilbert space should be normalizable.
Let ${\cal S}({\bf R}^3)$ be a set of smooth functions 
(${\bf R}^3 \mapsto {\bf C}$) 
with fast decrease.
Using a function $f \in {\cal S}({\bf R}^3)$, 
we define 
\begin{equation}
\hat \psi^\dagger(f) 
\equiv
\int f({\bf r}) \hat \psi^\dagger ({\bf r}) d^3 r
=
C_f \hat \Xi^\dagger + \hat \psi^{\prime \dagger}(f),
\end{equation}
where 
\begin{eqnarray}
C_f &\equiv& \int f({\bf r}) d^3 r,
\\
\hat \psi^{\prime \dagger}(f) 
&\equiv&
\int f({\bf r}) \hat \psi^{\prime \dagger}({\bf r}) d^3 r.
\end{eqnarray}
Using another function $g \in {\cal S}({\bf R}^3)$, 
we also define 
$\hat \psi^\dagger(g)$, $C_g$, and $\hat \psi^{\prime \dagger}(g)$
in a similar manner.
From these definitions and eq.\ (\ref{com_rel}), 
one can easily show that 
\begin{equation}
\langle N, \nu |
[\hat \Xi, \hat \Xi^\dagger]
| N', \nu' \rangle
, \
\langle N, \nu |
[\hat \Xi, \hat \psi'(f)]
| N', \nu' \rangle
, \
\langle N, \nu |
[\hat \Xi, \hat \psi^{\prime \dagger}(f)]
| N', \nu' \rangle
\ = s(1/V).
\label{com_rel_f}\end{equation}
Let us construct $\hat A(f)$ and $\hat B(g)$ from 
$\hat \psi(f)$ and $\hat \psi(g)$, respectively,
in a manner similar to eqs. (\ref{A}) and (\ref{B}).
When $f$ and $g$ are centered around ${\bf r}$ and ${\bf r'}$, 
respectively, 
we can show, using eq.\ (\ref{com_rel_f}), by calculations
similar to eqs.\ (\ref{tochuu}) and (\ref{calcAB}), that
\begin{equation}
\langle \alpha, {\rm G} | 
\hat A(f)
\hat B(g)
| \alpha, {\rm G} \rangle  
=
\langle \alpha, {\rm G} | 
\hat A(f)
| \alpha, {\rm G} \rangle  \langle \alpha, {\rm G} | 
\hat B(g)
| \alpha, {\rm G} \rangle  
+ s(1/V)
\quad
\mbox{
for $|{\bf r} - {\bf r'}| \sim V^{1/3}$}.
\label{AfBg}\end{equation}
The same relation can also be shown when operators $\hat A(f)$ 
and $\hat B(g)$ are constructed from smoothed operators of derivatives of
field operators.
It is therefore concluded that
the CSIB possesses the cluster property in the limit of $V \to \infty$
(while keeping the boson density constant).

%%%%% fragile states %%%%%%%%%

On the other hand, the NSIB does not possess the cluster property.
We can easily see this by taking
$\hat A({\bf r}) = \hat \psi^\dagger ({\bf r})$
and 
$\hat B({\bf r'}) = \hat \psi({\bf r'})$.
In fact, 
as $|{\bf r} - {\bf r'}| \sim V^{1/3}$,
\begin{eqnarray}
&&
\langle N, {\rm G} | 
\hat A({\bf r})
\hat B({\bf r'})
| N, {\rm G} \rangle  
-
\langle N, {\rm G} | 
\hat A({\bf r})
| N, {\rm G} \rangle  
\langle N, {\rm G} | 
\hat B({\bf r'})
| N, {\rm G} \rangle  
\nonumber \\
&& =
\langle N, {\rm G} | 
\{ \hat \Xi^\dagger + \hat \psi^{\prime \dagger} ({\bf r}) \}
\{ \hat \Xi + \hat \psi' ({\bf r'}) \}
| N, {\rm G} \rangle
-
\langle N, {\rm G} | 
\hat \psi^\dagger ({\bf r})
| N, {\rm G} \rangle
\langle N, {\rm G} | 
\hat \psi ({\bf r'})
| N, {\rm G} \rangle
\nonumber \\
&& =
\langle N, {\rm G} | 
\hat \Xi^\dagger \hat \Xi
| N, {\rm G} \rangle
+ s(1/V)
\nonumber \\
&& =
|\xi|^2 N
+ s(1/V),
\label{corrNSIB}\end{eqnarray}
which does not vanish as $V \to \infty$
(while keeping the boson density constant)
because $|\xi|^2 N = {\cal O}(1)$.

As another example, we examine the number-phase 
squeezed state of interacting bosons (NPIB) \cite{SI}.
Its energy density is almost degenerate 
(completely degenerate when $V \to \infty$) with the CSIB and NSIB, 
and
its wave function interpolates between these states.
The number (phase) uncertainty of the NPIB is larger (smaller) than 
that of the CSIB, hence the name `squeezed state'.
However, unlike the standard squeezed states,  
the NPIB has the minimum allowable value
of the number-phase uncertainty product, like the CSIB \cite{SI}.
The NPIB is characterized by two independent parameters $N$ and $\zeta$, 
and is defined by
\begin{equation}
|N,  \zeta, {\rm G} \rangle
\equiv
\sqrt{K(N,|\zeta|^2)}
e^{- |\zeta|^2/2}
\sum_{M=0}^{N}
\frac{\zeta^{* (N-M)}}{\sqrt{(N-M)!M!}}
|M, {\rm G} \rangle,
\label{NPIB}\end{equation}
where $K$ is a normalization constant \cite{SI}.
We henceforth assume that $N \gg |\zeta|^2 \gg 1$,
for which $K=1$,  
$\langle N \rangle = N - |\zeta|^2$, 
and
$\langle \delta N^2 \rangle = |\zeta|^2$
% (hence $\langle N \rangle/\langle \delta N^2 \rangle \gg 1$)
to a good approximation.\cite{SI}
After lengthy calculations, we can show that,
as $|{\bf r} - {\bf r'}| \sim V^{1/3}$,
\begin{eqnarray}
&&
\langle N,  \zeta, {\rm G} | 
\hat \psi^\dagger ({\bf r})
\hat \psi({\bf r'})
|N,  \zeta, {\rm G} \rangle  
-
\langle N, \zeta, {\rm G} | 
\hat \psi^\dagger ({\bf r})
|N,  \zeta, {\rm G} \rangle  
\langle N,  \zeta, {\rm G} | 
\hat \psi({\bf r'})
|N,  \zeta, {\rm G} \rangle  
\nonumber\\
&& =
%\frac{|\xi|^2 \langle N \rangle}{2\langle \delta N^2 \rangle}
|\xi|^2 \langle N \rangle/2\langle \delta N^2 \rangle
+ s(1/V)
\nonumber\\
&& \simeq
%\frac{|\xi|^2 N}{2 |\zeta|^2} + s(1/V)
|\xi|^2 N/2 |\zeta|^2 + s(1/V).
\label{corrNPIB}\end{eqnarray}
% As compared with the result for the NSIB, eq.\ (\ref{corrNSIB}), 
% we see that the correlation is reduced by a factor 
% $1/2 |\zeta|^2$.
Since $|\xi|^2 N = {\cal O}(1)$ and $|\zeta|^2 \gg 1$ (by assumption), 
this correlation is small.
However, 
it does not vanish as $V \to \infty$,
because $\zeta$ is basically independent of $N$ and $V$.
% when $\zeta$ is fixed independent of $V$.
Therefore, 
the NPIB does not possess the cluster property, either.
Comparing eq.\ (\ref{corrNPIB}) with eq.\ (\ref{corrNSIB}),
we see that the spatial correlation decreases as 
$|\zeta|^2$ ($=\langle \delta N^2 \rangle$) is increased, 
i.e, as the NPIB moves from near the NSIB toward the CSIB.
Note, however, that eq.\ (\ref{corrNPIB}) 
is applicable neither the NSIB nor CSIB 
(because $N \gg |\zeta|^2 \gg 1$ is assumed):
our main results are the {\em set} of 
eqs.\ (\ref{AB}), (\ref{corrNSIB}) and (\ref{corrNPIB}).

We have thus confirmed our conjecture that 
a robust state has the cluster property.
We must be careful in discussing the converse statement --- 
a state with the cluster property is robust --- for two reasons.
First, when a state has the cluster property at a particular time, 
it may evolve spontaneously into another state which does not have the cluster
property. %, and decohere quickly.
A typical example is the coherent state of free bosons as mentioned above.
Second, 
excited states generally have finite lifetimes, 
which tend to be shorter for higher-energy states.
This means that higher-energy states decohere quickly.
It would be non-trivial to define the robustness
for such states.
In this work, we have confined ourselves to the ground states.

In conclusion, we have examined the cluster property of 
various ground states of interacting many bosons confined 
in a box of a finite volume $V$.
It is shown that 
the robust ground state (CSIB) possesses the cluster property 
(in the limit of $V \to \infty$), whereas
the fragile ground states (NSIB and NPIB) do not.
Namely, 
the cluster property,
which is defined 
as a {\em static} property of a {\em closed} system,
% in the $V \to \infty$ limit,
is directly related to 
the robustness, which is 
a {\em dynamical} property of an {\em open} system. 
This fact, which was {\em assumed} previously as 
one of fundamental requirements which ensure consistency of  
a microscopic theory with macroscopic theories, 
has been shown to be satisfied by a realistic model 
of interacting particles for the first time.


\begin{thebibliography}{99}
\bibitem{eg}
Unfortunatelly, the complete list of all consistency conditions 
is not known yet.
\bibitem{odlro}
The cluster property should not be confused with the off-diagonal 
long-range order (ODLRO), which is defined by
$ % \begin{equation}
% \omega(\hat A({\bf r}) \hat B({\bf r'}))
\omega(
\hat \psi^\dagger ({\bf r}) \hat \psi^\dagger ({\bf r})
\cdots
\hat \psi ({\bf r'}) \hat \psi ({\bf r'})
\cdots
)
\not\to
0
$
as
$
| {\bf r} - {\bf r'} | \to \infty.
$ %\label{odlro}\end{equation}
For example, not only the CSIB but also the NSIB and NPIB 
have the ODLRO \cite{SI}, 
whereas the latter two states do not possess 
the cluster property [eqs.\ (\ref{corrNSIB}) and (\ref{corrNPIB})].
\bibitem{equal}
We here consider non-relativistic
theories, in which a space-like region is an equal-time hyper plane.
\bibitem{Ndep}
Strictly speaking, $\xi$ depends weakly on $N$.
Since we can show that this $N$ dependence does not alter our 
results, eqs. (\ref{AB}) and (\ref{AfBg}), we here describe the 
calculations in which this $N$ dependence is neglected.
\end{thebibliography}
\end{document}